
\magnification 1200
\baselineskip=17pt

\centerline{\bf NATURE OF DECOUPLING IN THE MIXED PHASE OF}
\bigskip
\centerline{\bf EXTREMELY TYPE-II LAYERED SUPERCONDUCTORS}
\vskip 50pt
\centerline{J. P. Rodriguez}
\medskip
\centerline{\it Dept. of Physics and Astronomy,
California State University, Los Angeles, CA 90032.}
\vskip 30pt
\centerline  {\bf  Abstract}
\vskip 8pt\noindent
The uniformly frustrated layered $XY$ model is analyzed
   in its Villain form. 
A decouple pancake
vortex liquid phase is identified.  
It  is bounded
by both first-order and second-order decoupling lines
in the magnetic field versus temperature plane.  	
These transitions,     respectively, can  account for
the flux-lattice
melting and for  the flux-lattice depinning 
observed in the mixed phase of clean high-temperature superconductors.

\bigskip
\noindent
PACS Indices:  74.60.-w, 74.25.Dw, 74.25.Ha, 74.60.Ge

\vfill\eject

The high-$T_c$ superconductor Bi$_2$Sr$_2$CaCu$_2$O$_8$
is perhaps the best known example of an extremely type-II 
layered superconductor.$^{1}$  
In the absence of bulk pinning, the vortex lattice that exists in
the mixed phase of this material melts across a first-order
line in the temperature-magnetic field 
plane for magnetic fields applied perpendicular to the layers.$^{2}$
The first-order line,
$H_{\perp} = H_m (T)$, begins at the zero-field critical point, $T_c$,
but it ends strangely  in the middle of the phase diagram.
The depinning       line
$T = T_{dp}(H_{\perp})$, which marks the point at which the flux lattice 
depins itself through   thermal excitations,
appears to be unrelated to this melting line.$^{3,4}$  
Considerable theoretical work has accompanied
such  observations.
For example, flux-lattice melting is observed in Monte Carlo (MC)
simulations of the frustrated $XY$ model,$^5$ as well as in simulations
of Ginzburg-Landau theory.$^{6}$  
The  Josephson coupling between layers 
is also predicted to diminish substantially 
at perpendicular fields, $B_{\perp}$,
that exceed the dimensional (2D-3D) cross-over scale,$^7$ 
$B_{\perp}^* \sim \Phi_0/\Lambda_0^2$.
Here, $\Lambda_0$ denotes the Josephson penetration length.
Last, it has been claimed that the layers decouple completely
through a first-order transition
at perpendicular field components $H_{\perp}$ many times larger than 
$B_{\perp}^*$.$^{8}$
A theoretical explanation of the critical endpoint of
the first-order melting line
in clean high-$T_c$ superconductors is lacking, however.

In this paper, we obtain a schematic 
phase diagram for the uniformly
frustrated $XY$ model composed of a  finite number
of weakly coupled layers,
which can describe
the mixed phase of extremely type-II layered superconductors.$^{5,9-12}$
In contrast to previous work,$^{7,8,12}$ the duality analysis$^{13-16}$ that
follows includes 
both gaussian and topological excitations$^{17-19}$
within the intra-layer vortex lattices.
We find a second-order melting line for
weakly coupled two-dimensional (2D) vortex lattices
in the limit of high perpendicular field.$^{17, 20}$  
It is then argued that this line
ends at an intersection with   a
first-order decoupling line$^8$ in the middle of 
the $T$-$H_{\perp}$ plane
(see Fig. 1).
The  theory potentially accounts for 
the same phenomenon that is observed in clean  
high-$T_c$ superconductors.$^{2,3}$
We also demonstrate how   the decoupling mechanism explicit 
in the duality analysis is intimately related to the entanglement
of flux lines perpendicular to the layers.  
This is consistent with the {\it supersolid} scenario,$^{19}$
with recent numerical  studies of the anisotropic $XY$ model with
frustration,$^{21}$ and with very recent experimental determinations 
of the phase diagram in clean high-temperature superconductors.$^{22}$
 
Consider the application  of a magnetic field 
to a layered superconductor in the extreme type-II limit,$^{1}$ 
$\lambda_L\rightarrow\infty$.
Magnetic screening effects are then  negligible and the theoretical
description
of the interior of the mixed phase reduces
to a   layered $XY$ model
in the presence of a uniform frustration.$^{5,9-12}$  In particular,
the thermodynamics
is determined by the kinetic energy 
$E_{XY} = -\sum_{r,\mu} J_{\mu} {\rm cos}
[\Delta_{\mu}\phi - A_{\mu}]$ as a function of the superconducting
phase $\phi(r)$.  Here,
$J_x = J_{\parallel}
= J_y$ and $J_z = J_{\perp}$
are the local phase rigidities, and
$A_{\mu} = (0, b_{\perp} x, -b_{\parallel}x)$ 
is the   vector potential. 
The magnetic induction 
parallel and perpendicular to the layers
is related to the frustration, $\vec b$, through the respective
identities $B_{\parallel} = (\Phi_0/2\pi d) b_{\parallel}$
and $B_{\perp} = (\Phi_0/2\pi a) b_{\perp}$.
Here $a$ denotes the square lattice constant, which is of order
the zero-temperature coherence length, while
$d$ represents the spacing in between consecutive layers.  
Consider now  the corresponding partition function,
$Z[p]=\int {\cal D} \phi\,  e^{-E_{XY}/k_B T} e^{ i\sum p  \phi}$, 
in the (dual)  Villain form,$^{13-16}$ 
which reads
$$Z[p] = \sum_{\{n_{\mu}(r)\}}\Pi_{r}
\delta\Biggl[\sum_{\nu}\Delta_{\nu} n_{\nu}|_{r} - p(r)\Biggr]
{\rm exp}\Biggl[-\sum_r\Biggl({1\over {2\beta_{\parallel}}}
\vec n^2 + {1\over{2\beta_{\perp}}} n_z^2
+ i\sum_{\nu} n_{\nu}A_{\nu}\Biggr)\Biggr].\eqno (1)$$
Above, $n_{\mu}(r)$ is  an integer link-field
on the layered lattice structure  of
points $r=(\vec r, l)$, with   $\mu = \hat x, \hat y, \hat z$ and
$\vec n=(n_x,n_y)$.  Also, we set  $\beta_{\parallel,
\perp}=J_{\parallel,\perp}/k_B T$. 
Now decompose the parallel field $\vec n$ into 
transverse and longitudinal 
parts $\vec n(\vec r, l) = \vec n\,^{\prime}(\vec r, l)
- \vec n_-(\vec r, l) + \vec n_-(\vec r, l-1)$,
where the transverse and longitudinal fields,
$\vec n^{\prime}$ and $\vec n_-$, respectively
satisfy the constraints
$\vec\nabla\cdot\vec n^{\prime} = 0$ and
$\vec\nabla\cdot\vec n_- = n_z$ (for $p = 0$), 
with $\vec\nabla = (\Delta_x, \Delta_y)$.$^{12,23}$
Take next  the potential
representation $\vec n_{-}=-\vec \nabla\Phi$
for the longitudinal inter-layer field.
This  yields the expression
$\Phi(\vec r, l) =
\sum_{\vec r\,^{\prime}} G^{(2)}(\vec r - \vec r\,^{\prime})
n_z(\vec r\,^{\prime}, l)$ for the potential, where 
$G^{(2)} = -\nabla^{-2}$
is  the Greens function for the square lattice.
We then obtain the form
$Z[0] = Z_{\rm CG}\cdot \Pi_{l} Z_{\rm DG}[0]$ for
the partition function (1), 
where 
$$\eqalignno{
Z_{\rm CG} =  \sum_{\{ n_z\}} y_0^{N[n_z]} {\rm exp}
\Biggl[  -{1\over{2\beta_{\parallel}}}\sum_{l}
\sum_{\vec r_1, \vec r_2}  q_l(\vec r_1)
G^{(2)}(\vec r_1 - \vec r_2)
q_l(\vec r_2)&
-i\sum_{l}\sum_{\vec r} n_z(\vec r, l) A_z(\vec r)\Biggr]\times\cr
&\times
\Pi_l (Z_{\rm DG}^{\prime}[q_l]/Z_{\rm DG}[0])& (2) \cr}$$
is    an  inter-layer Coulomb gas (CG) factor.
Here,  $q_l(\vec r) = n_z(\vec r, l-1) - n_z(\vec r, l)$  is
the fluxon charge that collects onto layer $l$ and
$y_0 = {\rm exp}(-1/2\beta_{\perp})$ is the fugacity
that is raised to the power
$N[n_z] = \sum_{\vec r,l} n_z^2(\vec r,l)$ per configuration.$^{16}$
Note that a fluxon, $n_z(\vec r,l) = \pm 1$, represents a
vortex ring that lies in between layers $l$ and $l+1$ at the
point $\vec r$.$^{24}$
The remaining factors
$$\eqalignno{
Z_{\rm DG}^{\prime}[q_l] = \sum_{\{\vec n\}}&\Pi_{\vec r}
\delta[\vec\nabla\cdot\vec n|_{\vec r, l}
- q_l(\vec r)]\times\cr
&\times {\rm exp}\Biggl[-{1\over {2\beta_{\parallel}}}
\sum_{\vec r} \vec n^2(\vec r,l)
-i\sum_{\vec r} \vec n(\vec r,l)
\cdot\vec A^{\prime}(\vec r)\Biggr] C_{\rm sw}[q_l]  & (3)\cr}$$
above represent modified 
2D $XY$ models with uniform frustration 
corresponding to each layer,
with a modified in-plane vector potential
$\vec A^{\prime} = \vec A + 
i\beta_{\parallel}^{-1}[\vec n_{-}(\vec r,l-1) - \vec n_{-}(\vec r,l)]$
in the gauge $\vec\nabla\cdot\vec A = 0$,
and with an extra    weight factor
$C_{\rm sw}[q_l] = {\rm exp} \{-\sum_{\vec r}
(2\beta_{\parallel})^{-1}[\vec n_{-}(\vec r,l-1) - \vec n_{-}(\vec r,l)]^2\}$.
In physical terms, the CG factor (2) describes the Josephson coupling
between layers, whereas the discrete gaussian
(DG) model factors (3) describe the 
thermodynamics of  pancake vortices within each layer.
The last factor in expression (2), however,
represents the renormalization
of the Josephson coupling
due to misalignments of pancake vortices between
layers.$^{7,8}$  This important correction was omitted
without proper justification
in previous work.$^{12,23-25}$

Consider now the weak-coupling limit, 
$y_0\rightarrow 0$, in which case the $n_z$ fluxons are dilute.$^{24}$ 
The modified 2D $XY$ model (3)
can then be analyzed in the continuum limit, in
which case we obtain the identification
$$Z_{\rm DG}^{\prime}[q_l]/Z_{\rm DG}[0]
= \Bigl\langle {\rm exp} \Bigl[  i \sum_{\vec r} 
q_l(\vec r) \phi_{\rm vx}(\vec r,l)\Bigr]\Bigr\rangle_{J_{\perp} = 0}
\eqno (4)$$
with the vortex component of the phase correlations within an isolated
layer $l$:$^{13,14}$
i.e.,  $\nabla^2 \phi_{\rm vx}  =  0$.
It is instructive to consider a single  neutral  pair of unit $n_z$ charges
that lie  in between
layers $l^{\prime}$ and $l^{\prime}+1$,
separated by $\vec r$. 
The renormalization to the Josephson
coupling that is  encoded in the last factor of expression (2) 
then becomes the gauge-invariant product
$\Pi_{l}
(Z_{\rm DG}^{\prime}[q_l]/Z_{\rm DG}[0]) =
C_{l^{\prime}}^{\prime} (\vec r) C_{l^{\prime}+1}^{\prime *}(\vec r)$ 
of the corresponding  phase autocorrelation functions,
$$C_{l}^{\prime} (\vec r) = \Bigl\langle {\rm exp}
\Bigl[i\phi_{\rm vx}(0,l) - i\phi_{\rm vx}(\vec r,l) 
\Bigr]\Bigr\rangle_{J_{\perp} = 0}\ ,
\eqno (5)$$
within isolated layers.
In the asymptotic
limit, $\vec r\rightarrow \infty$, this function has
a magnitude of the form
$|C_l^{\prime}(\vec r)| = g_0 (r_0/r_0^{\prime})^{\eta_{\rm sw}}
(r_0/|\vec r\,|)^{\eta_{\rm vx}}$ for
$|\vec r|\ll \xi_{\rm vx}$, and a magnitude   of the form
$|C_l^{\prime}(\vec r)| = 
g_0 (|\vec r\,|/r_0^{\prime})^{\eta_{\rm sw}}\, {\rm exp} 
(-|\vec r\,|/\xi_{\rm vx})$ for 
$|\vec r|\gg \xi_{\rm vx}$.$^{13-15}$  
Here, $\eta_{\rm sw} = (2\pi\beta_{\parallel})^{-1}$
and  $\eta_{\rm vx}$ are, respectively, 
the spin-wave and the vortex components
of        the correlation exponent inside layer $l$, while
$\xi_{\rm vx}$ is the corresponding
phase correlation length.  Also, the length
$r_0 = a_{\rm vx} / 2^{3/2} e^{\gamma_E}$ is  set by
the inter-vortex scale,
$a_{\rm vx} = (\Phi_0/B_{\perp})^{1/2}$, and by   
Euler's constant,$^{14}$ $\gamma_E$,
while $r_0^{\prime} = a/2^{3/2} e^{\gamma_E}$.
We therefore obtain the
effective layered CG ensemble
$$\eqalignno{ Z_{\rm CG} \cong  \sum_{\{n_z\}}  y^{N[n_z]}{\rm exp}
\Biggl\{  - {1\over 2}\sum_{l}
\sum_{\vec r_1 \neq \vec r_2}^{\qquad\prime} q_l(\vec r_1)
\Bigl[ \eta_{2D}
 & {\rm ln} (r_0 / |\vec r_1 - \vec r_2|)
 - V^{[q_l]}_{\rm string} (\vec r_1,  \vec r_2)\Bigr]
q_l(\vec r_2) \cr
&-i\sum_{l}\sum_{\vec r}^{\qquad\prime} n_z(\vec r, l) A_z(\vec r)\Biggr\}
 &(6)\cr}$$
that has been coarse grained up to the natural
ultraviolet  scale $a_{\rm vx}$.
In particular, the sums above are restricted to a square sublattice
with lattice constant $a_{\rm vx}$. 
This  requires the introduction of an effective coarse-grained fugacity 
$y = g_0  (a_{\rm vx}/a)^{2} y_0$.
At relatively small separation
$|\vec r_1 - \vec r_2| \ll \xi_{\rm vx}$, the fluxons experience
a pure Coulomb interaction ($V_{\rm string}^{[q_l]} = 0$) 
set by the 2D correlation exponent
$\eta_{2D} = \eta_{\rm sw} + \eta_{\rm vx}$.
At large separations $|\vec r_1 - \vec r_2|\gg\xi_{\rm vx}$,
on the other hand,
the fluxons experience
a pure ($\eta_{2D} = 0$) confining interaction
$V_{\rm string}^{[q_l]} (\vec r_1,\vec r_2) = 
  |\vec r_1  -  \vec r_2|/\xi_{\rm vx}$
between those  points      $\vec r_1$ and $\vec r_2$
in layer $l$
that are connected by a  string [see Eq. (3) and ref. 26].

We shall now see that
the  effective layered CG ensemble (6)
can be employed to determine   the macroscopic nature of the 
Josephson effect   in
the weak-coupling limit, $y\rightarrow 0$.
Let us first identify the macroscopic  intra-layer phase rigidity,$^{27}$
 $\bar J_{\parallel} = k_B T / 2\pi \eta_{2D}$.   
The above Coulomb gas ensemble indicates that
fluxons ($n_z = \pm 1$) are in a plasma state at temperatures
below the naive decoupling temperature
$k_B T_* = 4\pi \bar J_{\parallel}$
when  quasi        long-range intra-layer
phase correlations are present,$^{23,24}$ $\xi_{\rm vx} = \infty$
and $V^{[q_l]}_{\rm string} = 0$,
while that they form a
dilute gas of bound neutral pairs of size $\xi_{\rm vx}$
at high temperatures
when short-range intra-layer phase correlations exist, $\xi_{\rm vx} < \infty$.
We now quote the expression for the perpendicular phase
rigidity of the $XY$ model (2) in terms of fluxons:$^{15}$
$$\rho_s^{\perp}
= {\cal N}^{-1}
\Bigl \langle \Bigl [\sum_{\vec r, l} n_z (\vec r, l)\Bigr]^2 \Bigr\rangle
  k_B T ,\eqno (7)$$
where $\cal N$ denotes the number of links between layers.
It is obtained directly from the duality transformation (1) and
from the definition 
$\rho_s^{\perp} = {\partial^2\over{\partial A_z^2}} (G_{\rm cond}/{\cal N})$
for this quantity.   Under periodic boundary conditions,
the low-temperature  (plasma) phase thus sustains a macroscopic
Josephson effect ($\rho_s^{\perp}\neq 0$) due the presence of free fluxons,
whereas the high-temperature 
(dielectric) phase does not due to the absence
of free fluxons ($\sum n_z = 0$).  
Since the ordering temperature of a single layer is typically
much less than $T_*$ (see below), we conclude that
the only thermodynamic phases that are possible
at weak coupling
are  a coupled superconductor at low temperatures
and a decoupled ``normal''
state at high temperatures.
This indicates that  neither
the Friedel scenario$^{24,28}$ (decoupled superconducting
layers) nor the line-liquid state$^{9, 12, 23}$ (coupled normal layers)
are likely to be  thermodynamic states  in the absence of disorder.

It is also important to determine the size   of the local Josephson coupling. 
Consider the macroscopically decoupled phase
at weak coupling,
where short-range intra-layer
phase correlations are present: $\xi_{\rm vx} < \infty$.
Inter-layer fluxon ($n_z$) pairs are then
bound by a confining string.
Comparison of the CG ensemble (6) with the layered $XY$ model (1) 
yields Koshelev's formula$^{10}$
$$\langle e^{i\phi_{l,l+1}}\rangle \cong  y_0
\int d^2 r 
C_l(\vec r) C_{l+1}^* (\vec r) e^{  - i b_{\parallel} x} /a^2\eqno (8)$$
for the local Josephson coupling 
(see refs. 15 and 16).
Here, $\phi_{l,l+1}(\vec r)
= \phi (\vec r,l+1)- \phi (\vec r,l)-A_z(\vec r)$
is the gauge-invariant phase difference between consecutive layers,
while $C_l(\vec r)$ is the 
phase autocorrelator for layer
$l$ in isolation [i.e., replace $\phi_{\rm vx}\rightarrow \phi$
in Eq. (5)].   Eq. (8) implies that
$\partial^2 \,{\rm ln}\, Z[0]  /
 \partial \beta_{\perp} \partial B_{\parallel} = 0$ at
$B_{\parallel} = 0$.
We then  have  a null line tension 
for Josephson vortices, since
$\varepsilon_{\parallel} = 0 = \varepsilon_{\parallel}|_{J_{\perp} = 0}$.
In conclusion, we recover the previous result
(7) that macroscopic Josephson
coupling is absent 
in the weak-coupling limit
if intra-layer
phase correlations are short range.  
Next,  observe that scaling considerations
yield the form
$\int d^2 r C_l C_{l+1}^*  = 
f_0 \xi_{\rm vx}^2$
for the integral on the  right-hand side of Eq. (8), 
where $f_0$ is of order unity.$^{10}$
Substitution into Eq. (8) then yields 
Koshelev's formula
$$\langle {\rm cos}\, \phi_{l,l+1}\rangle \cong
 f_0 (\xi_{\rm vx}/a)^2  y_0\eqno (9)$$
for the local Josephson coupling in such case
(see ref. 16).
[Note that scaling implies
the functional form $\xi_{\rm vx} =  a_{\rm vx}  e(\beta_{\parallel})$
for  the 2D correlation length .] 

We  can  now analyze the $XY$ model (1) composed of
a  finite number of weakly-coupled layers
with uniform frustration, $B_{\perp}$.
Consider first the weak-coupling  limit,
 $\langle {\rm cos}\, \phi_{l,l+1}\rangle\rightarrow 0$,
which by Eq.  (9)   is reached at infinitely high
perpendicular fields. 
It is well known that an isolated lattice of 2D vortices (3) melts 
at a temperature $k_B T_m^{(2D)} \cong J_{\parallel}/20$,
above which quasi long-range positional correlation
in the vortex lattice
is lost.$^{20}$  The transition is driven by the unbinding of
dislocation pairs and it  is expected theoretically to
be second-order.$^{17}$  
We shall now make the plausible assumption that the nature of phase
coherence in the 2D vortex lattice
 is locked to the nature of the positional correlations,
such that $\xi_{\rm vx}$ diverges exponentially as the temperature
approaches $T_m^{(2D)}$ in the disordered phase.
By the previous  analysis, we then conclude that the layers
show a macrsoscopic Josephson effect  at low temperature $T < T_m^{(2D)}$ 
signalled by a positive phase rigidity between layers (7), 
while that they are decoupled 
($\rho_s^{\perp} = 0$)
at high temperature $T > T_m^{(2D)}$.
The former low-temperature phase is best described  by
2D vortex lattices that display a Josephson effect.
The decoupled high-temperature phase, on the other hand,
corresponds to a liquid of intra-layer
``pancake'' vortices.$^{7,8}$

Consider next the weak-coupling regime,
$\langle {\rm cos}\, \phi_{l,l+1}\rangle \ll 1$, at   high
perpendicular fields 
$B_{\perp} \gg  B_{\perp}^*$ [see Eq. (9)].
Eq. (9) indicates that the selective high-temperature expansion
breaks down
($\langle {\rm cos}\,\phi_{l,l+1}\rangle >    1$)
in the decoupled phase at a temperature $T_{\times}$
set roughly  by the identification of length scales
$\Lambda_0\sim\xi_{\rm vx}(T_{\times})$.
We now observe that the layered $XY$ model (1) without
frustration can also be described by the Coulomb gas ensemble    (6),
but with the natural ultraviolet length scale replaced
globally by $a_{\rm vx}\rightarrow a$.
By analogy with what is
presently understood for the layered $XY$ model without
frustration,$^{29}$
we conclude that a   second-order transition should
take place in the weak-coupling regime
at a temperature $T_m$ that lies
inside of  the dimensional crossover window
$T_m^{(2D)} < T < T_{\times}$.

And      what happens as the local Josephson coupling (9)
approaches unity,
which can be achieved by lowering the perpendicular field?
The CG ensemble (6) is screened in the low-temperature Josephson-coupled
phase, $T <  T_m^{(2D)}$, for small effective fugacity.
This implies that no phase transition is possible as a function of
field there.$^{13,14}$  Nevertheless, Eq.  (9) clearly indicates that the
selective high-temperature expansion, $y_0\rightarrow 0$,
breaks down at perpendicular fields below $B_{\perp}^*$.
In such case, a crossover into a flux-line lattice regime must therefore
take place.$^7$
At high temperatures $T > T_{\times}$, on the other hand,
the CG ensemble (6) is  confining for small fugacity, $y\ll 1$.  
In particular, the string interaction (6) 
binds together  dilute fluxon-antifluxon pairs into  stable
dipoles of dimension $\xi_{\rm vx}$.
In the limit of dense
fluxons, $y\rightarrow 1$, these  dipoles  disassociate, however.
This is due to the ineffectiveness  of the string when        
the distance, $r_s$, between neighboring dipoles  
is small in comparison to the length of the string, $\xi_{\rm vx}$.
The  system must therefore experience a (inverted)
{\it first-order} phase transition into a screened CG
above a critical coupling 
 $\langle {\rm cos}\, \phi_{l,l+1} \rangle_D$, since there is no  
diverging length scale.
Monte Carlo simulation of the layered $XY$ model with
uniform  frustration indicates that
$\langle {\rm cos}\, \phi_{l,l+1}\rangle_D$
is constant and of order unity.$^{11}$
By Eq. (9), we therefore
expect a first-order decoupling transition at a   perpendicular field 
$$H_{D} = (\bar f_0\Phi_0/a^2)
\Bigl({1\over 2} \beta_{\perp}
/\langle {\rm cos}\, \phi_{l,l+1}\rangle_D\Bigr)
 \eqno (10)$$
of order $\beta_{\parallel} B_{\perp}^*$
for high temperatures, $T > T_{\times}$.
This results from the  replacement
$y_0\rightarrow {1\over 2}\beta_{\perp}$ (see ref. 16).
The  phenomenology$^{30}$ 
$J_{\perp} = E_{J0}  (T_c-T)/T_c$
for the Josephson  energy 
in the vicinity of the zero-field transition at $T_c$ yields   
the dependence$^{2,7,8}$ 
$H_{D} (T) =   \gamma_2^{-2} H_{c2} (T)$
for the ``cosine'' $XY$ model, where 
$H_{c2} (T)\sim (\Phi_0/a^2) (T_c - T)/T$ 
is the mean-field perpendicular
upper-critical field, and where
$\gamma_2\sim 
(\langle e^{i\phi_{l,l+1}}\rangle_D/f_0)^{1/2}\cdot (k_B T_c/E_{J0})^{1/2}$
is an  effective anisotropy parameter.
Similar results for the first-order decoupling transition (10)
were obtained previously
using the elastic medium description of vortex
matter in layered superconductors.$^{7,8}$
The above discussion is summarized
by the schematic phase diagram in  Fig. 1.  

We shall now give a physical interpretation of the results just obtained 
from the duality analysis.  Consider two parallel vortices
along the $z$-axis that exchange positions in between layers $l$ and
$l+1$.  A moment's thought determines that the exchange creates a
(fluxon) vortex loop that lies in between those layers.
Fluxon charge$^{24}$ ($n_z$) can therefore {\it entangle} vortex lines
aligned perpendicular to the layers.  We conclude that entanglement
is what actually drives the decoupling transitions shown in Fig. 1.
This picture is consistent with ({\it a}) the classification of the
coupled 2D vortex-lattice phase as a type of super-solid matter,$^{19}$
with ({\it b}) recent Monte Carlo simulations of the anisotropic $XY$
model with frustration that conclude  that entanglement is what drives the
first-order melting transition of the vortex lattice,$^{21}$ and with
({\it c}) recent experimental work that also finds entanglement to
be what controls the location of the  critical endpoint
of the vortex-lattice melting transition
in clean high-temperature superconductors.$^{22}$
This web of facts strongly supports the results obtained here.

We shall close by comparing the present theory for the uniformly frustrated
layered $XY$ model with known experimental results for the 
mixed phase of Bi$_2$Sr$_2$CaCu$_2$O$_8$,
which is   extremely type-II and  layered.
This system shows  first-order melting in the absence of 
bulk pinning.
In agreement with Fig. 1, the melting
line ends in the middle of the phase diagram.$^2$
This suggest 
identifying it with  the
decoupling transition at $H_{\perp} = H_{D}$.
Also, the depinning       line, $T = T_{dp}$,
is nearly vertical in such
materials for perpendicular fields above $B_{\perp}^*$.$^{3,4}$
This  suggests identifying
$T_{dp}$ with $T_m$ in Fig. 1.$^{31}$
Finally, although Monte Carlo simulations of the frustrated $XY$ model with 
anisotropy  do observe a unique first-order vortex-lattice
melting transition, no indication of the critical endpoint predicted here
has been reported.$^{11}$  Eq. (10) implies, however, that the critical
anisotropy  parameter, 
$\gamma^{\prime} = (J_{\parallel}/J_{\perp})^{1/2}$,
is high: e.g., $\gamma_c^{\prime}  \sim 33$ for perpendicular
fields $B_{\perp} = \Phi_0/56 a^2$ at $T_m^{(2D)}$.  
The condition that the
Josephson penetration length, $\Lambda_0 = \gamma^{\prime}_c a$,
be smaller than $L/2\pi$, where $L$ is the linear dimension of each layer,
then indicates that extensive MC simulations$^{21}$
are necessary in order to see
the critical endpoint predicted here.$^{15}$

The author is grateful for the hospitality of the
Instituto de Ciencia de Materiales de Madrid, where
part of this work was completed,
and to Tony Gonzalez-Arroyo for discussions.
He also thanks D. Huse for bringing ref. 19 to his
attention.
This work was supported in part by National
Science Foundation grant \# DMR-9322427.

\vfill\eject
\centerline{\bf References}
\vskip 16 pt

\item {1.} M. Tinkham, Physica C {\bf 235}, 3 (1994).

\item {2.} E. Zeldov, D. Majer, M. Konczykowski,
V.B. Geshkenbein, V.M. Vinokur, and H. Shtrikman,
Nature {\bf 375}, 373 (1995).

\item {3.} D.T. Fuchs, E. Zeldov, T. Tamegai, S. Ooi, M. Rappaport
and H. Shtrikman, Phys. Rev. Lett. {\bf 80}, 4971 (1998);
B.  Khaykovich, E. Zeldov, D. Majer, T.W. Li, P.H. Kes, and
M. Konczykowski, Phys. Rev. Lett. {\bf 76}, 2555 (1996).

\item {4.} J.H. Cho, M.P. Maley, S. Fleshler, A. Lacerda and L.N. Bulaevskii,
Phys. Rev. B {\bf 50}, 6493 (1994).

\item {5.} R.E. Hetzel, A. Sudb\o, D.A. Huse, Phys. Rev. Lett. {\bf 69},
518 (1992).

\item {6.} R. Sasik and D. Stroud, Phys. Rev. Lett. {\bf 75}, 2582 (1995);
J. Hu and A.H. MacDonald, Phys. Rev. B {\bf 56}, 2788 (1997).

\item {7.} L.I. Glazman and A.E. Koshelev, Phys. Rev. B {\bf 43}, 2835 (1991).

\item {8.} L.L. Daemen, L.N. Bulaevskii, M.P. Maley and
J.Y. Coulter,  Phys. Rev. B {\bf 47}, 11291 (1993).

\item {9.} Y.H. Li and S. Teitel, Phys. Rev. B {\bf 47}, 359
(1993); {\it ibid} {\bf 49}, 4136 (1994).

\item {10.} A.E. Koshelev, Phys. Rev. Lett. {\bf 77}, 3901 (1996).

\item {11.} A.E. Koshelev, Phys. Rev. B {\bf 56}, 11201 (1997).

\item {12.} J.P. Rodriguez, J. Phys. Cond. Matter {\bf 9}, 5117 (1997). 

\item {13.} J.V. Jos\' e, L.P. Kadanoff, S. Kirkpatrick and
D.R. Nelson, Phys. Rev. B {\bf 16}, 1217 (1977).

\item {14.} C. Itzykson and J.  Drouffe, {\it Statistical Field Theory}, 
vol. 1, (Cambridge Univ.  Press, Cambridge, 1991) chap. 4.

\item {15.} J.P. Rodriguez, Phys. Rev. B {\bf 62}, 9117 (2000).

\item {16.} The results obtained here   are recovered
for the case of the conventional ``cosine'' $XY$
model through the replacement 
$y_0\rightarrow {1\over 2}\beta_{\perp}$
of the fugacity (see ref. 15).

\item {17.} J.M. Kosterlitz and D.J. Thouless,
J. Phys. C {\bf 6}, 1181 (1973).

\item {18.} M. Feigel'man, V.B. Geshkenbein, and A.I. Larkin, 
Physica C {\bf 167}, 177 (1990). 

\item {19.} E. Frey, D.R. Nelson, and D.S. Fisher, 
Phys. Rev. B {\bf 49}, 9723 (1994). 

\item {20.} S.A. Hattel and J.M. Wheatley,
Phys. Rev. B {\bf 51}, 11951 (1995).

\item {21.} Y. Nonomura, X. Hu and M. Tachiki, Phys. Rev. B {\bf 59},
R11657 (1999).

\item {22.} G.W. Crabtree, W.K. Kwok, L.M. Paulius,
A.M. Petrean, R.J. Olsson,
G. Karapetrov, V. Tobos and W.G. Moulton, Physica C {\bf 332},
71 (2000).

\item {23.} J.P. Rodriguez, Europhys.  Lett. {\bf 31}, 479 (1995).

\item {24.}  S.E. Korshunov, Europhys. Lett. {\bf 11}, 757 (1990).

\item {25.} J.P. Rodriguez, Europhys. Lett. {\bf 39}, 195 (1997); 
Europhys. Lett. {\bf 47}, 745 (E) (1999);
 Phys. Rev. B {\bf 58}, 944 (1998).

\item {26.} A. Polyakov, Phys. Lett. {\bf 72} B, 477 (1978).

\item {27.} D.R. Nelson and J.M. Kosterlitz, Phys. Rev. Lett. {\bf 39},
1201 (1977).


\item {28.} J. Friedel, J. Phys. (Paris) {\bf 49}, 1561 (1988).

\item {29.} S. Hikami and T. Tsuneto, Prog. Theor. Phys. {\bf 63}, 387 (1980);
see also S.R. Shenoy and B. Chattopadhyay, Phys. Rev. B {\bf 51}, 9129
(1995).

\item {30.} V. Ambegaokar and A. Baratoff, Phys. Rev. Lett. {\bf 10}, 486
(1963); {\bf 11}, 104 (E) (1963).

\item {31.} The combination of a rigid vortex-lattice with
surface barriers is assumed to be the pinning mechanism.


\vfill\eject
\centerline{\bf Figure Captions}
\vskip 20pt
\item {Fig. 1.}   Shown is a schematic phase diagram for the uniformly
frustrated $XY$ model (1) made up of a finite number of weakly-coupled
layers at $B_{\parallel} = 0$.  
A vestige of the vertical second-order line may   extend down to
lower fields in the form of a cross-over 
(compare with ref. 3).  The Josephson temperature
$T_{\rm J} = J_{\perp}/k_B$  is assumed to be smaller
than the scale of the figure.  Notice that the inequality
$T_{\rm J} < T_m^{(2D)}$  required by the phase diagram indicates
a minimum anisotropy parameter 
$\gamma^{\prime} = (J_{\parallel} / J_{\perp})^{1/2}$  
equal to about four, since 
$k_B T_m^{(2D)} \cong J_{\parallel} / 20$.

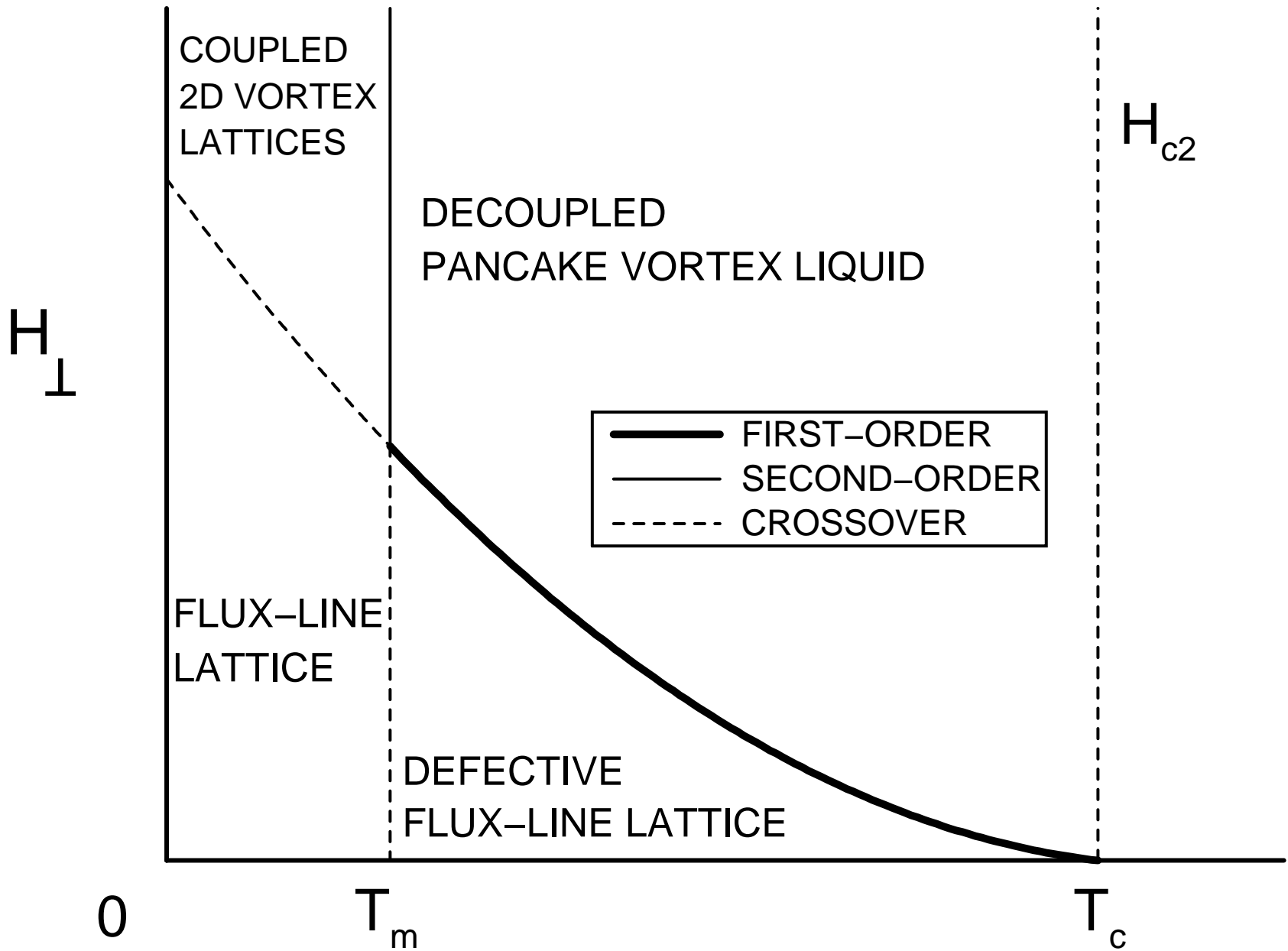
\end